# Quantum Interference of Stored Coherent Spin-wave Excitations in a Two-channel Memory

Hai Wang*,[1], Shujing Li[1], Zhongxiao Xu[1], Xingbo Zhao[1], Lijun Zhang[1], Jiahua Li[1], Yuelong Wu, Changde Xie[1], Kunchi Peng[1], and Min Xiao[1,2]

[1] *The State Key Laboratory of Quantum Optics and Quantum Optics Devices, Institute of Opto-Electronics, Shanxi University, Taiyuan, 030006, People's Republic of China*

[2] *Department of Physics, University of Arkansas, Fayetteville, Arkansas, 72701, USA*

**Quantum memories are essential elements in long-distance quantum networks[1] and quantum computation[2]. Significant advances have been achieved in demonstrating relative long-lived single-channel memory at single-photon level in cold atomic media[3-6]. However, the qubit memory corresponding to store two-channel spin-wave excitations (SWEs) still faces challenges, including the limitations resulting from Larmor procession[7-8], fluctuating ambient magnetic field[9-11], and manipulation/measurement of the relative phase between the two channels[12-15]. Here, we demonstrate a two-channel memory scheme in an ideal tripod atomic system, in which the total readout signal exhibits either constructive or destructive interference when the two-channel SWEs are retrieved by two reading beams with a controllable relative phase. Experimental result indicates quantum coherence between the stored SWEs. Based on such phase-sensitive storage/retrieval scheme, measurements of the relative phase between the two SWEs and Rabi oscillation, as well as elimination of the collapse and revival of the readout signal, are experimentally demonstrated.**

In recent years, by using the magnetic-field-insensitive atomic clock transitions and reducing the decay rates of atomic coherence, quantum memory time at single-photon level has been increased from microseconds to milliseconds time scale[3-4] and the optical coherent memory time has reached to even one second time scale[5-6]. To store an arbitrary polarization state[7] or generate entanglement between photons and atomic ensembles[8,10,11], an atomic qubit memory, i.e. encoding its two basis states in two coherent spin-wave excitations (SWEs), is necessary. Unlike the single-channel quantum memory, in the qubit memory one has to consider the relative phase between the two stored SWEs. Such relative phase will experience a time evolution[8] when a finite magnetic field (which is applied to define the quantization axis for the system) is present. Besides, to implement quantum information processing based on the atomic qubit memory, necessary manipulation and measurements related to the relative phase of the qubit will have to be realized[12-15]. Since different Zeeman coherences have different Larmor procession periods, the retrieved signal from such quantum memory will experience a "collapse and revival" phenomenon, which has been observed[16]. In the earlier experiments for quantum memory[9-11], the applied magnetic fields are typically very weak (~10 mG) and the achieved quantum memory times are up to ~10 μs. In the case of such weak magnetic field, the quantization axis for determining the interactions between photon polarization and the internal atomic states is not well defined due to the magnetic-field fluctuations in different directions, which can make qubit states out of order and cause decoherence. Recently, storage of an arbitrary polarization state[7], as well as entanglement of a photon and a SWE[8], were investigated with a stronger magnetic field to better define the quantization axis. The former experiment clearly shows that the stored photon polarization state is deteriorated by the fast Larmor procession[7], and the later experiment indicates that the stored qubit signals can only be read out at certain times due to Larmor procession[8], which significantly limit the applications of such quantum memory devices.

Two-channel optical storage in a four-level inverted-Y[17] or tripod[18] atomic configuration provides an ideal system to study the qubit memory. In such systems, the optical signal can be stored simultaneously in two adjacent atomic coherences (i.e. SWEs), as shown for the tripod system in Fig.1a. An earlier experiment with one writing/reading beam and two signal beams in a tripod system has shown resonance beating in the readout signal due to the lifted Zeeman sub-level degeneracy of the ground states in a magnetic field[19].

In this Letter, we demonstrate a two-channel memory by employing an ideal four-level tripod system, as shown in Fig.1a, with $^{87}$Rb atoms in a MOT (see Methods). The weak signal beam (P) can be simultaneously stored in the two coherent SWEs $\tilde{\sigma}_{ca}$ and $\tilde{\sigma}_{ba}$ by two circularly-polarized writing beams (W$^+$, W$^-$). When reading by two circularly-polarized beams (R$^+$, R$^-$) with a certain relative phase $\delta_R = \varphi_R^+ - \varphi_R^-$, quantum interference is induced by the light-matter interaction between the two stored coherent collective SWEs, which can show a total maximum or minimum readout of the stored information depending on the relative phase between the two reading light beams $\delta_R$. By varying $\delta_R$, the total readout signal shows a sinusoidal interference pattern. The two-channel storage scheme with the phase manipulation in the readout process has several advantages over the previously demonstrated optical memories. First, with phase-sensitive quantum interference between the two-channel SWEs, the relative phase of the two SWEs and the population projected into a given superposition of the two SWEs can be experimentally measured. Second, the current experiment shows that the collapses and revivals of the readout signal due to Larmor procession in the well-defined tripod system can be eliminated by tuning the relative phase between the two reading beams. Third, although the current experiment is done with a weak coherent signal, the scheme and experimental system can be extended by using a single-photon source as the input probe signal, so experiments for long-lived qubit memory, qubit gate operation, and Hong-Ou-Mandel interference[18] can be performed.

Before showing the experimental results, we first present a simple theoretical description using the generalized dark-state polariton concept[17], which is modified to fit the current experimental situation. The generalized dark-state polaritons, including a quantum field $\hat{\varepsilon}_p(z,t)$ and a total collective atomic spin wave $\hat{S}(z,t)$, can be written as:

$$\hat{\Psi}(z,t) = \cos\theta(t) \cdot \hat{\varepsilon}_p(z,t) - \sin\theta(t) \cdot \sqrt{N}\hat{S}(z,t) \quad , \qquad (1)$$

where the total collective atomic spin wave has the form

$$\hat{S}(z,t) = \frac{\Omega_C^+(t)}{\sqrt{|\Omega_C^+(t)|^2 + |\Omega_C^-(t)|^2}} \tilde{\sigma}_{ca}(z,t) + \frac{\Omega_C^-(t)}{\sqrt{|\Omega_C^+(t)|^2 + |\Omega_C^-(t)|^2}} \tilde{\sigma}_{ba}(z,t) . \qquad (2)$$

The mixing angle is given by $tg\theta(t) = \frac{g\sqrt{N}}{\sqrt{|\Omega_C^+(t)|^2 + |\Omega_C^-(t)|^2}}$; $\Omega_C^+(t)$ ($\Omega_C^-(t)$) is the Rabi frequency of right- (left-) circularly-polarized writing or reading coupling beam. When the two writing beams (assuming with equal amplitude) are adiabatically turned off at the same time, the signal beam (P) is then simultaneously stored in the two coherent SWEs $\tilde{\sigma}_{ca}$ and $\tilde{\sigma}_{ba}$, with the relative phase information ($\delta_w = \varphi_w^+ - \varphi_w^-$) imprinted onto them (See Methods). During a storage time τ, the atomic coherences $\tilde{\sigma}_{ba}(z,t_1)$ and $\tilde{\sigma}_{ca}(z,t_1)$ will experience Larmor procession with a procession frequency $\Omega_L$. After time τ, the two coherent reading beams, with a relative phase $\delta_R = \varphi_R^+ - \varphi_R^-$, are turned on and they collinearly pass through the atomic medium. For a zero total phase difference, e.g. $\Delta = \delta_R - \delta_w + 2\Omega_L\tau = 0$, the relative phase experienced by the two spin-wave components $\tilde{\sigma}_{ca}$ and $\tilde{\sigma}_{ba}$ in $\hat{S}(z,t)$ (see Eq.(4) in Methods) is in phase, so it gives a maximum value due to constructive interference, which maps onto the readout optical signal. When the total phase difference $\Delta = \delta_R - \delta_w + 2\Omega_L\tau = \pi$, the relative phase experienced by the two spin-wave components $\tilde{\sigma}_{ca}$ and $\tilde{\sigma}_{ba}$ in $\hat{S}(z,t)$ is out of phase, so it reaches to a minimum value due to destructive quantum interference, which reflects at the output optical signal as a low readout intensity (see Methods).

The experiment is carried out in a standard MOT with $^{87}$Rb atoms. The atoms are prepared in an ideal tripod system by selectively pumping the atomic population into the $5S_{1/2}$, F=1, m=+1 state[20]. The details of the MOT, time sequence, and state preparation procedure are described in the Methods section. The experimentally prepared tripod system can be well described by the simple energy diagram shown in Fig.1a and the configurations of the writing, reading, and signal beams are shown in Fig.1b with all beams split from a single high-power diode laser. The relative phase difference is introduced by separating the horizontal and vertical components of the reading beam and controlled by using the electro-optical modulator (EOM).

Figure 2a shows the storage and retrieved signal from a single (left) channel (blocking the right writing/reading beams), which corresponds to the typical Λ-type system. Following the standard procedure for optical storage and retrieval[21] (see Methods), the retrieved signal is read after a 380 ns time delay (right peak). The right channel alone shows the same behavior. When both the left and right writing beams are turned on and have the same power ($\left|\Omega_W^+(t)\right| = \left|\Omega_W^-(t)\right|$) as in the single-channel case, the optical signal is stored simultaneously into two coherent SWEs ($\tilde{\sigma}_{ca}$ and $\tilde{\sigma}_{ba}$). If one reading beam (either $R^+$ or $R^-$) is used, the readout signal is about the same as in the single-channel case. Figures 2b and 2c present the readout signals from individual $\tilde{\sigma}_{ba}$ and $\tilde{\sigma}_{ca}$ (with only one reading beam on), respectively. When both reading beams ($R^+$ and $R^-$) are on at the same time and the total phase is tuned to satisfy $\Delta = 0$ by adjusting the relative phase between the two reading beams ($\delta_R$), a maximal readout signal is obtained (Fig.2d). The retrieved signal intensity is twice of the single channel case at the same storage time of 380 ns, as indicated in the calculation (see Methods, Eq.(6)). These results clearly show the quantum interference between the two stored collective SWEs mediated by the coherent reading light beams.

When the total phase difference $\Delta$ is adjusted to be $\pi$ by tuning $\delta_R$, no optical signal is retrieved from the atomic medium even though both reading beams are turned on at 380 ns. After a delay time τ of 3.4 μs, the atomic medium is read again with other (writing) laser beams, as shown in Fig.3a. For comparison, Fig.3b presents the readout result when the stored SWEs were not perturbed before the system is read at 3.4 μs storage time with the horizontally-polarized writing beam (which is composed of $\sigma^+$- and $\sigma^-$- polarized components with a fixed relative phase of $\pi/2$). These results indicate that although the stored SWEs have been read with two reading beams at 380 ns for $\Delta = \pi$, they have not been converted into the optical signal due to destructive quantum interference between them and remained in the atomic medium, until they are read again with other laser beams at a later time.

Figure 4 (square points) shows the total retrieved signal as a function of $\delta_R$ at τ=380 ns.

For fixed $\delta_W = \pi/2$ and Larmor procession (e.g. $2\Omega_L \tau$ =constant), $\delta_R$ is proportional to $\Delta$. The retrieval efficiency shows a sinusoidal interference pattern as $\delta_R$ is varied (Eq. (6) in Methods). At the same time, this relation can be viewed as the population projected into the given state $\hat{\sigma}_+^{\delta_R}|a\rangle$, and the interference pattern shows that the population in the $\hat{\sigma}_+^{\delta_R}|a\rangle$ state depends on the relative phase $\delta_R$ between the two basis states (see Methods). After the stored collective SWEs are first read by a pair of reading beams (at 380 ns), the atomic medium is read again by the horizontally-polarized (writing) laser beam at 3.4 μs. The retrieved signals at 3.4 μs are also shown in Fig.4 (circular points) as a function of $\delta_R$ (for the first pair of reading beams). It is clear that if all the stored SWEs in $\hat{S}(z,t)$ are mapped into signal field completely at the earlier time (380 ns), the reading by the second beam will not be able to retrieve any information. However, if the first reading at t=380 ns did not retrieve information from the SWEs (due to $\Delta=0$), the second reading at 3.4 μs will have maximum readout signal.

In the tripod atomic system with a small magnetic field (~300 mG), the Zeeman sublevels $|b\rangle$ and $|c\rangle$ are split, which causes Larmor precession for the stored collective SWEs[14]. With the phase difference between the two reading beams optimized for readout at 380 ns storage time, the total readout signal intensity as a function of storage time shows a sinusoidal behavior, as shown in Fig.5 (circular points). Actually, such sinusoidal curve evolving with storage time reveals the population Rabi oscillation between the superposition states $\hat{\sigma}_+|a\rangle$ and $\hat{\sigma}_-|a\rangle$ (where $\hat{\sigma}_\pm|a\rangle = (\hat{\sigma}_{ba} \pm \hat{\sigma}_{ca})|a\rangle/\sqrt{2}$, see Methods for details), which may be applied to rotation operations of qubit memory in quantum information processing[14-15]. However, such oscillatory behavior in the readout intensity can severely limit the ability to retrieve arbitrarily polarized states[7-8], as well as to probe long-lived entanglement between photons and SWEs at any desired times. In the current phase-sensitive storage/retrieval scheme we demonstrate that by adjusting the relative phase difference between the two reading beams, such "collapse and revival" behavior in readout signal intensity due to Larmor procession can be completely compensated (i.e. $\delta_R$ can compensate $2\Omega_L\tau$ in total phase difference $\Delta$), as shown in Fig.5 (square points). In this case the maximal readout signal can be accessed at any storage time and controlled by adjusting the relative phase difference between the two reading beams.

The ability of preparing this ideal four-level tripod system[20] for two-channel memory in the readout process allows us to isolate contributions of the interested coherent atomic states and to demonstrate the quantum interference between $\hat{\sigma}_{ba}$ and $\hat{\sigma}_{ca}$. It also enables us to measure the population projected into a given coherence superposition state, Rabi oscillation between the superposition of $\hat{\sigma}_+|a\rangle$ and $\hat{\sigma}_-|a\rangle$ states, and compensation of Larmor procession in the readout process. In the single tripod system, we have experimentally shown that if an optical signal is stored in only one coherent SWE (say $\tilde{\sigma}_{ba}$) and read at later times from another coherent SWE ($\tilde{\sigma}_{ca}$), no signal can be retrieved. This shows that there is no direct energy exchange between the two stored collective SWEs, which can be very useful for certain quantum information processing applications.

The presented two-channel storage scheme and the coherent phase-sensitive readout technique have significant advantages in extending the single-channel memory systems for quantum information processing, especially for qubit memory, and have overcome some important obstacles in employing atomic ensembles as optical memory devices. This two-channel storage/retrieval scheme provides a unique way to measure the relative phase of the qubit memory and probe the population in a given coherent superposition state. The accessed Rabi oscillation (with a lifetime of ~90 μs) between two superposition states of the SWEs can have potential applications in performing qubit gate operations. Also, by compensating the Larmor procession via tuning the relative phase between the two reading beams, the collapses and revivals of the total readout signal are eliminated and the total readout signal keeps at its maximum for more than 100 μs. Such phase compensation method can also be applied to the experiments of the quantum memory for photon polarization states[7] and entanglement of photons and atomic ensembles[8]. In those experiments, if one could first measure the evolution of the relative phase of the qubit memory due to Larmor procession, and alter the relative phase between the two retrieved orthogonally-polarized photons by an EOM according to the measured phase data, the changes of photon polarization states due to Larmor procession[7,8] could then be compensated, and therefore the memory time would be significantly increased.

**Methods**

**Two-channel dark-state polaritons.** According to Fig.1a, the quantized signal field $\hat{E}_p^+(z,t) = \hat{\varepsilon}_p(z,t)e^{-i\omega_p t + ik_p z}$ is $\sigma^-$-polarized, and couples to the transition from level |a> to level |e>, where $\hat{\varepsilon}_p(z,t)$ is the slowly-varying amplitude with an initial phase of zero. The writing or reading coupling field consists of two polarization components in the atomic medium, i.e. $\sigma^+$-polarized $\tilde{E}_C^+(z,t) = |E_C^+(z,t)|e^{-i\omega_C t + ik_C z - i\varphi_C^+}$ and $\sigma^-$-polarized $\tilde{E}_C^-(z,t) = |E_C^-(z,t)|e^{-i\omega_C t + ik_C z - i\varphi_C^-}$, with slowly-varying Rabi frequencies defined as $\Omega_C^+(t) = |\Omega_C^+(t)|e^{-i\varphi_C^+}$ ($|\Omega_C^+(t)| = \mu|E_C^+(t)|/\hbar$) and $\Omega_C^-(t) = |\Omega_C^-(t)|e^{-i\varphi_C^-}$ ($|\Omega_C^-(t)| = \mu|E_C^-(t)|/\hbar$), driving the transitions from |c> to |e> and |b> to |e> in the four-level tripod system, respectively.

In the storage process, the writing beam (including two components $\Omega_W^+(t)$ and $\Omega_W^-(t)$) is adiabatically switched off over the time interval [$t_0$, $t_1$] as the probe pulse enters into the cold atomic assemble, so the probe field $\hat{\varepsilon}_P^{in}(z,t)$ is mapped onto the total collective atomic spin wave $\hat{S}_w(z,t)$ and stored in the cold atomic assemble. Under the condition of $|\Omega_w^+(t)| = |\Omega_W^-(t)|$, this total collective atomic spin wave $\hat{S}_w(z,t)$ can be written, from Eq (2) in the text, as[17]

$$\hat{S}_w(z,t_1) = \frac{1}{\sqrt{2}} \tilde{\sigma}_{ca}(z,t_1) e^{-i\varphi_w^+} + \frac{1}{\sqrt{2}} \tilde{\sigma}_{ba}(z,t_1) e^{-i\varphi_w^-}, \quad (1)$$

where $\tilde{\sigma}_{ca}(z,t_1)$ and $\tilde{\sigma}_{ba}(z,t_1)$ are two collective SWEs and are defined as

$$\tilde{\sigma}_{\beta\alpha}(z,t) = \frac{1}{N_z} \sum_{j=1}^{N_j} \tilde{\sigma}_{\beta\alpha}^j(t) = F(z,t)\hat{\sigma}_{\beta\alpha} e^{i\omega_{\alpha\beta}(z-ct)}. \quad \hat{\sigma}_{\beta\alpha} = |\beta\rangle\langle\alpha| \text{ is spin flip operator}$$

($\alpha = a$, $\beta = b$ or c) and $F(z,t)$ is the normalized distribution function for the SWE, which is related to $\hat{\varepsilon}_P^{in}(z,t)$. According to the dark-state polariton theory[17,21] in the case of $\theta \approx \pi/2$ (see Eq.(1) in the text), the signal field $\hat{\varepsilon}_P^{in}(z,t)$ is mapped into the total collective atomic spin wave $\hat{S}_w(z,t)$, and the calculated two SWEs are:

$$\tilde{\sigma}_{ba}(z,t_1) \propto \frac{1}{\sqrt{2}} \varepsilon_P^{in}(z-z_{01}, t_0) \hat{\sigma}_{ba} e^{i\varphi_w^-}, \quad (2a)$$

$$\tilde{\sigma}_{ca}(z,t_1) \propto \frac{1}{\sqrt{2}} \varepsilon_P^{in}(z-z_{01}, t_0) \hat{\sigma}_{ca} e^{i\varphi_w^+}, \quad (2b)$$

where $\varepsilon_P^{in}(z-z_{01}, t_0)$ is the slowly-varying amplitude of the input probe field at $z-z_{01}$ ($z_{01} = \int_{t_0}^{t_1} dt' v_g(t')$) and time $t=t_0$. In a finite magnetic field the atomic coherences (SWEs) $\tilde{\sigma}_{ba}(z,t_1)$ and $\tilde{\sigma}_{ca}(z,t_1)$ experience a Larmor procession, and after a storage time interval of $\tau = t_2 - t_1$, they then become

$$\tilde{\sigma}_{ba}(z,t_2) \propto \frac{1}{\sqrt{2}} \varepsilon_P^{in}(z-z_{01}, t_0) \hat{\sigma}_{ba} e^{i\varphi_w^- + i2\Omega_L \tau}, \quad (3a)$$

$$\tilde{\sigma}_{ca}(z,t_2) \propto \frac{1}{\sqrt{2}} \varepsilon_P^{in}(z-z_{01}, t_0) \hat{\sigma}_{ca} e^{i\varphi_w^+}. \quad (3b)$$

$\Omega_L = g_F \mu_B B / \hbar$ is the Larmor procession frequency and $2\Omega_L$ equals to the frequency shift between the Zeeman sublevels $|c\rangle$ ($|F=2, m=-1\rangle$) and $|b\rangle$ ($|F=2, m=+1\rangle$) in the magnetic field B. At the time $t_2$, if two reading beams, $|\Omega_R^+(t)| e^{-i\varphi_R^+}$ and $|\Omega_R^-(t)| e^{-i\varphi_R^-}$ (with $|\Omega_R^+(t)| = |\Omega_R^-(t)|$), are used to read the two-channel SWEs stored in the tripod atomic medium, the total collective atomic spin wave $\hat{S}(z,t)$ becomes:

$$\begin{aligned}\tilde{S}_R(z,t_2) &= \frac{1}{\sqrt{2}}\tilde{\sigma}_{ca}(z,t_2)e^{-i\varphi_R^+} + \frac{1}{\sqrt{2}}\tilde{\sigma}_{ba}(z,t_2)e^{-i\varphi_R^-} \\ &= \frac{1}{2}\varepsilon_p^{in}(z-z_{01},t_0)\left(\hat{\sigma}_{ca}e^{i\varphi_w^+ - i\varphi_R^+} + \hat{\sigma}_{ba}e^{i\varphi_w^- - i\varphi_R^- + i2\Omega_L\tau}\right)\end{aligned}, \quad (4)$$

which will be turned into the released total optical signal. The total readout signal can then be written as[17,21]

$$\varepsilon_P^{out}(z,t) \propto \langle \hat{S}_R(z-z_{02},t_2)\rangle \propto \frac{1}{2}\varepsilon_P^{in}(z-z_0,t_0)(e^{-i\delta_R + i\delta_W - i2\Omega_L\tau} + 1), \quad (5)$$

where $z_{02} = \int_{t_2}^{t} v_g(t')dt'$, $z_0 = z_{01} + z_{02}$, $\delta_R = \varphi_R^+ - \varphi_R^-$, and $\delta_w = \varphi_w^+ - \varphi_w^-$.

The individual signal $\varepsilon_{P+}^{out}$ or $\varepsilon_{P-}^{out}$ retrieved separately from the two memory channels ($\tilde{\sigma}_{ca}$ and $\tilde{\sigma}_{ba}$) can be easily calculated through Eq.(4) and Eq.(5) under the condition of $\Omega_R^-(t) = 0$ or $\Omega_R^+(t) = 0$ with $|\varepsilon_{P+}^{out}(z,t)| = |\varepsilon_{P-}^{out}(z,t)| = \varepsilon_{single}^{out}(z,t) \propto \frac{1}{\sqrt{2}}\varepsilon_P^{in}(z-z_0,t_0)$. The total retrieved signal intensity can be expressed as:

$$|\varepsilon_P^{out}|^2 \propto 2|\varepsilon_{single}^{out}|^2 \cos^2(\frac{\delta_R}{2} - \frac{\delta_W}{2} + \Omega_L\tau), \quad (6)$$

where $\varepsilon_{single}^{out}$ denotes the amplitude of the readout signal retrieved from a single channel (only with $\Omega_R^+(t)$ or $\Omega_R^-(t)$). From Eq.(6), it is easily seen that when $\Delta = \delta_R - \delta_W + 2\Omega_L\tau = 0$ and $\pi$, $|\varepsilon_P^{out}|^2 \approx 2|\varepsilon_{single}^{out}|^2$ and $|\varepsilon_P^{out}|^2 = 0$, respectively, which show the constructive and destructive interferences at the total optical readout signal.

The total retrieval efficiency of this two-channel storage system is expressed by $R_e = \int |\varepsilon_P^{out}|^2 dt \Big/ \int |\varepsilon_P^{in}|^2 dt$, where $|\varepsilon_P^{in}|$ is the amplitude of the input signal. Substituting Eq.(3) into this expression and considering the decoherence of the atomic memory, the total retrieval efficiency as a function of storage time is given by $R_e = A[1 + \cos(2\Omega_L t - \phi)]e^{-t/t_0}$, where $t_0$ is the lifetime, and $A = \int |\varepsilon_{single}^{out}|^2 dt \Big/ \int |\varepsilon_P^{in}|^2 dt$ stands for the total retrieval efficiency as the

storage time approaches to zero, which mainly depends on the optical depth of the atomic medium[22].

Next, we calculate the time evolution of the relative phase between the two SWEs and show how such time evolution can result in a population Rabi oscillation of the superposition states. Let's consider an atom excited into the SWEs and assume that the SWEs $\hat{\sigma}_{ba}$ and $\hat{\sigma}_{ca}$ have the same probability amplitude ($1/\sqrt{2}$) with an initial relative phase $\delta_w$. During the storage time τ, the relative phase between $\hat{\sigma}_{ba}$ and $\hat{\sigma}_{ca}$ will experience a time evolution due to Larmor procession. In the representation with two basic excitation states $\hat{\sigma}_{ba}|a\rangle = |b\rangle$ and $\hat{\sigma}_{ca}|a\rangle = |c\rangle$, the total atomic SWE state $|\psi(t)\rangle$ can be written as:

$$|\psi(t)\rangle = (e^{i2\Omega_L t + i\varphi_w^-}\hat{\sigma}_{ba} + e^{i\varphi_w^+}\hat{\sigma}_{ca})|a\rangle/\sqrt{2}$$
$$= e^{i\Omega_L \tau + i(\frac{\varphi_w^+}{2} + \frac{\varphi_w^-}{2})}\left(e^{i\Omega_L t - i\frac{\delta_w}{2}}\hat{\sigma}_{ba} + e^{-i\Omega_L t + i\frac{\delta_w}{2}}\hat{\sigma}_{ca}\right)|a\rangle/\sqrt{2} \quad (7)$$

Using another two basic states $\hat{\sigma}_+|a\rangle$ and $\hat{\sigma}_-|a\rangle$ ($\hat{\sigma}_+|a\rangle = (\hat{\sigma}_{ba} + \hat{\sigma}_{ca})|a\rangle/\sqrt{2}$ and $\hat{\sigma}_-|a\rangle = (\hat{\sigma}_{ba} - \hat{\sigma}_{ca})|a\rangle/\sqrt{2}$), the collective SWE state $|\psi(t)\rangle$ can be expressed as

$$|\psi(t)\rangle = e^{i\Omega_L t + i(\frac{\varphi_w^+}{2} + \frac{\varphi_w^-}{2})}\left(\cos\left(\Omega_L t - \frac{\delta_W}{2}\right)\hat{\sigma}_+ + i\sin\left(\Omega_L t - \frac{\delta_W}{2}\right)\hat{\sigma}_-\right)|a\rangle \quad . \quad (8)$$

This expression indicates that the population experiences a Rabi oscillation between the states $\hat{\sigma}_+|a\rangle$ and $\hat{\sigma}_-|a\rangle$ with a frequency $\Omega_L$, which is due to the interaction between a finite d.c. magnetic field and multi-level atoms.

Writing the expression of the collective SWE state $|\psi(t)\rangle$ in the representation with the two basic states $\hat{\sigma}_+^{\delta_R}|a\rangle = (e^{-i\delta_R/2}\hat{\sigma}_{ba} + e^{i\delta_R/2}\hat{\sigma}_{ca})|a\rangle/\sqrt{2}$ and $\hat{\sigma}_-^{\delta_R}|a\rangle = (e^{-i\delta_R/2}\hat{\sigma}_{ba} - e^{i\delta_R/2}\hat{\sigma}_{ca})|a\rangle/\sqrt{2}$, we can obtain the population in the basis state $\hat{\sigma}_+^{\delta_R}|a\rangle$ for an excited atom as

$$P = \cos^2\left(\frac{\delta_R}{2} - \frac{\delta_W}{2} + \Omega_L t\right) \quad . \quad (9)$$

Considering the distribution for the SWE $F(z,t)$, the total population in the state $\hat{\sigma}_+^{\delta_R}|a\rangle$ for the excited atoms can be calculated to be

$$P_{total} \propto 2\left|\varepsilon_{sigle}^{out}\right|^2 \cos^2\left(\frac{\delta_R}{2} - \frac{\delta_W}{2} + \Omega_L t\right) = \left|\varepsilon_P^{out}\right|^2 \quad . \tag{10}$$

This means that the total readout signal read by the two reading beams with a relative phase $\delta_R$ is just the population projected into the state $\hat{\sigma}_+^{\delta_R}|a\rangle$.

**Experimental details.** In the experiment, more than $10^9$ $^{87}$Rb atoms are trapped in a standard MOT (with an atomic cloud of diameter ~5mm). The measured optical depth is about 1.5 for the transition from $|5S_{1/2}, F=1\rangle$ to $|5P_{1/2}, F'=1\rangle$ and the trap temperature can reach ~200 μK within ~990 ms. Then, the MOT (including cooling and repumping lasers, as well as the trapping magnetic field) is switched off, and, at same time, a bias magnetic field of 300 mG along the z axis (see Fig.1) is applied, so the z-direction quantization axis is well defined. Then a 780 nm, right-circularly-polarized pumping laser beam, coupling to the transition from $|5S^{1/2}, F=1\rangle$ to $|5^2P_{3/2}, F'=1\rangle$, and the writing laser beam (with two circularly-polarized components in the atomic medium) are turned on, so most of the atoms (>90%) are prepared into the $|5^2S_{1/2} F=1, m=+1\rangle$ state[21]. After 300 μs, the probe pulse (with a pulse length of 100 ns) enters into the atomic medium. Successively, the pumping and writing laser beams are quickly blocked, thus partial signal pulse is adiabatically mapped onto the cold atoms as two collective spin-wave excitations in the tripod system. After waiting for a time duration of 380 ns, the reading beams ($R^-$ and/or $R^+$ with a controllable relative phase between them by the EOM) are turned on to read the stored collective spin-wave excitations. We have also read the atomic spin-wave excitations after a time delay of 3.4 μs in certain cases to test the phase-sensitive readout process in this two-channel memory system.

**The Relative phase measurement.** The relative phase ($\delta_R = \varphi_R^+ - \varphi_R^-$) of the left- and right-circularly-polarized reading beams is measured by using a homodyne detection setup, which is shown in Fig. 1b. By using an EOM in the two reading beams, a relative phase ($\delta_R$) is introduced between the s- and p-polarized reading beams. Going through a λ/4 waveplate whose axis is set to be in the direction with a $45^0$ angle to the p direction, the s- and p-polarized reading beams become the left- (σ⁻-) and right-(σ⁺-) circularly-polarized beams, respectively. The σ⁻- and σ⁺- reading beams then propagate through the cold atoms and a polarization beam splitters PBS3 (whose axis is in the direction with a $45^0$ angle from the p direction). The transmitted beam from PBS3 consists of the left- and right-circularly-polarized components of the reading beams, which are detected by APD2. By changing the voltage applied to the EOM, a sinusoidal interference signal is detected, which can be used to determine the relative phase $\delta_R$. The relative phase ($\delta_w = \varphi_w^+ - \varphi_w^-$) between the left- and right-circularly-polarized writing beam components is fixed at $\pi/2$ since the p-polarized writing beam has a $45^0$ angle respect to the axis of the λ/4 waveplate.

**References**

1. Duan, L.-M., Lukin, M. D., Cirac, J. I. & Zoller, P., Long-distance quantum communication with atomic ensembles and linear optics. *Nature* **414**, 413 (2001).
2. Lvovsky, A. I., Sanders, B. C. & Tittel, W., Optical quantum memory. *Nature Photonics* **3**, 706 (2009).
3. Zhao R., Dudin, Y. O., Jenkins, S. D., Campbell, C. J., Matsukevich, D. N., Kennedy, T. A. B. & Kuzmich, A., Long-lived quantum memory. *Nature Phys*. **5**, 100 (2009).
4. Zhao, B., Chen, Y.A., Bao, X.H., Strassel, T., Chuu, C.S., Jin, X.M., Schmiedmayer, J., Yuan, Z.-S., Chen, S. & Pan, J.-W., A millisecond quantum memory for scalable quantum networks. *Nature Phys*. **5**, 95 (2009).
5. Schnorrberger, U., Thompson, J. D., Trotzky, S., Pugatch, R., Davidson, N., Kuhr, S. & Bloch, I., Electromagnetically induced transparency and light storage in an atomic Mott insulator. *Phys. Rev. Lett.* **103**, 033003 (2009)
6. Zhang, R., Garner, S. R. & Hau, L. V., Creation of long-term coherent optical memory via controlled nonlinear interactions in Bose-Einstein condensates. *Phys. Rev. Lett.* **103**, 233602 (2009).
7. Tanji, H., Ghosh, S., Simon, J., Bloom, B. & Vuletić, V., Heralded single-magnon quantum memory for photon polarization states. *Phys. Rev. Lett.* **103**, 043601 (2009)
8. Dudin, Y. O., Jenkins, S. D., Zhao, R., Matsukevich, D. N., Kuzmich, A. & Kennedy, T. A. B., Entanglement of a Photon and an Optical Lattice Spin Wave. *Phys. Rev. Lett.* **103**, 020505 (2009).
9. Chanelière, T., Matsukevich, D. N., Jenkins, S. D., Lan, S.-Y., Kennedy, T. A. B. & Kuzmich, A., Storage and retrieval of single photons transmitted between remote quantum memories, *Nature* **438**, 833 (2005).

10. Matsukevich, D. N., Chanelière, T., Bhattacharya, M., Lan, S.-Y., Jenkins, S. D., Kennedy, T. A. B. & Kuzmich, A., Entanglement of a photon and a collective atomic excitation. *Phys. Rev. Lett.* **95**, 040405 (2005).

11. Choi, K. S., Deng, H., Laurat, J.& Kimble, H. J., Mapping photonic entanglement into and out of a quantum memory. *Nature* **452**, 67 (2008).

12. Duan, L.-M., Entangling many atomic ensembles through laser manipulation. *Phys. Rev. Lett.* **88**, 170402 (2002).

13. Duan, L.-M. & Raussendorf, R., Efficient quantum computation with probabilistic quantum gates. *Phys. Rev. Lett.* **95**, 080503 (2005).

14. Langer, C., Ozeri, R., Jost, J. D., Chiaverini, J., DeMarco, B., Ben-Kish, A., Blakestad, R. B., Britton, J., Hume, D. B., Itano, W. M., Leibfried, D., Reichle, R., Rosenband, T., Schaetz, T., Schmidt, P. O. & Wineland, D. J., Long-lived qubit memory using atomic ions. *Phys. Rev. Lett.* **95**, 060502 (2005).

15. Moehring, D. L., Maunz, P., Olmschenk, S., Younge, K. C., Matsukevich, D. N., Duan, L.-M., & Monroe, C., Entanglement of single-atom quantum bits at a distance. *Nature* **449**, 68 (2007).

16. Matsukevich, D. N., Chanelière, T., Jenkins, S. D., Lan, S.-Y., Kennedy, T. A. B. & Kuzmich, A., Observation of dark state polariton collapses and revivals. *Phys. Rev. Lett.* **96**, 033601 (2006).

17. Joshi, A. & Xiao, M., Generalized dark-state polaritons for photon memory in multilevel atomic media. *Phys. Rev. A* **71**, 041801(R) (2005).

18. Raczyński, A., Zaremba, J. & Zielińska-Kaniasty, S., Beam splitting and Hong-Ou-Mandel interference for stored light. *Phys. Rev. A* **75**, 013810 (2007).

19. Karpa, L., Vewinger, F. & Weitz, M., Resonance beating of light stored using atomic spinor polaritons. *Phys. Rev. Lett.* **101**, 170406 (2008).

20. Wang, B., Han, Y.-X., Xiao, J.-T., Yang, X.-D., Zhang, C.-H., Wang, H., Xiao, M. & Peng, K.-C., Preparation and determination of spin-polarized states in multi-Zeeman- sublevel atoms. *Phys. Rev. A* **75,** 051801(R) (2007).

21. Phillips, D. F., Fleischhauer, A., Mair, A., Walsworth, R. L. and Lukin, M. D., Storage of light in atomic vapor, Phys. Rev. Lett. **86**, 783 (2001).

22. Gorshkov, A. V., Andre, A., Fleischhauer, M., Sorensen, A. S. & Lukin, M. D., Universal approach to optimal photon storage in atomic media. Phys. Rev. Lett. **98**, 123601 (2007).

**Acknowledgement:** We acknowledge funding support from the National Natural Science Foundation of China (No.10874106, 60736040, 60821004, 10904086), and the 973 Program (2010CB923103). M.X. acknowledges partial support from US NSF.

*Correspondence and requests for materials should be addressed to H.W. (e-mail: wanghai@sxu.edu.cn)

**Figure Captions:**

**Figure 1│ Schematic view and experimental setup. a,** Four-level tripod system in the D1 line of $^{87}$Rb atom. Ground level |a> is the $\left|5S_{1/2}, F=1, m=+1\right\rangle$ state; |b> and |c> are the $\left|5S_{1/2}, F=2, m=+1\right\rangle$ and $\left|5S_{1/2}, F=2, m=-1\right\rangle$ states, respectively; and |e> is the $\left|5P_{1/2}, F'=1, m=0\right\rangle$ state. The dotted (blue) line W$^+$(R$^+$) is the right-circularly-polarized left writing (reading) beam. The solid (red) line W$^-$(R$^-$) is the left-circularly-polarized right writing (reading) beam. The dashed (green) line P is the left-circularly-polarized probe beam. $\sigma_{ba}$ ($\sigma_{ca}$) indicates the induced atomic coherence (spin-wave excitation) between states |a> and |b> (|a> and |c>). **b,** All the laser beams are split from the same grating feedback diode laser. The linearly-polarized writing beam W (about 1.2 mW) provides the left- and right-circularly-polarized W$^-$ and W$^+$ components in the MOT (with a 300 mG magnetic field in the z direction); The phase difference between the two orthogonally-polarized reading beams is provided by the EOM. These two linearly-polarized beams are converted into the left- and right-circularly-polarized reading beams (about 1.4 mW, each), respectively, by the λ/4 waveplate. BD1 and BD2 are two beam displacing polarizers. The probe beam (about 55 μW) is left-circularly-polarized and is shifted 6.8 GHz from the laser beam by two high-frequency AOMs, which goes through the cold atomic sample with a small angle (∼0.4°) from the writing/reading beams. PBS1-PBS3: polarization cube beam splitters; APD1 and APD2: avalanche photo detectors.

**Figure 2│ Retrieval efficiency for different storage/retrieval configurations. a,** One-channel storage and retrieval for a three-level Λ-type system as a reference. **b,** Mapping the signal field *P*

simultaneously into two collective spin-wave excitation channels, but read the stored atomic coherence with only one (left-circularly-polarized R⁻) beam. In this case, only stored coherent spin-wave excitation $\sigma_{ba}$ is retrieved at 380 ns. **c,** Optical information is stored in both channels, but only one (left) channel ($\sigma_{ca}$) is read. **d,** Optical signal is simultaneously stored in both channels by adiabatically turning off the *p*- polarized writing beam, and then the two reading beams are turned on simultaneously after a storage time of 380 ns. By adjusting the relative phase of the two reading beams, the total spin wave *S(z,t)* has a zero total phase difference ($\Delta \approx 0$), and thus the total readout signal is enhanced due to quantum interference between the two spin-wave excitations. The total phase difference $\Delta = \delta_R - \delta_w + 2\Omega_L \tau$ is determined as following: the relative phase for the two reading beams is measured to be $\delta_R \approx 0.2\pi$ by using optical interference (see Methods), the *p*-polarized writing beam has a relative phase of $\delta_w = \pi/2$ for its right- and left-components, the Larmor procession $2\Omega_L \tau \approx 0.3\pi$ for the Larmor frequency $\Omega_L \approx 2\pi \times 0.21 MHz$ and the storage time $\tau \approx 380 ns$, so $\Delta \approx 0$.

**Figure 3│ Non-retrievability with out-of-phase reading beams. a,** The optical signal is stored simultaneously into two channels, and first read with a pair of out-of-phase reading beams ($\Delta \approx \pi$) at 380 ns after storage. No signal is retrieved due to destructive interference in the two collective spin-wave excitations. At 3.4 μs after storage, the atomic medium is read again with the original (*p*-polarized) writing beam W (which has two circularly-polarized components with a relative phase of $\pi/2$), and the remain collective spin-wave excitations are then retrieved. **b,** As a reference, the system has not been read at 380 ns after storage, but only read at 3.4 μs by the original writing beam. Note that the retrieved signal at 3.4 μs after storage is obviously much smaller than that at 380 ns. Since the mean storage time is measured to be about 90 s, the reduction in readout signal intensity at 3.4 s is mainly due to the unoptimized total phase of the total atomic spin wave $\hat{S}(z,t)$ (i.e. $\Delta \approx 2\pi + 0.6\pi$) due to Larmor procession, when the *p*-polarized writing beam reads the memory at 3.4μs.

**Figure 4│ Retrieval efficiency or population in the state $\hat{\sigma}_+^{\delta_R}|a\rangle$ as a function of total**

**phase difference** $\delta_R$. The blue square points are the total retrieved signal intensities at 380 ns storage time for different total phase differences. The solid blue line is a best fit to the data. The red circular points are the retrieved signal intensities at 3.4 μs storage time, after the system has been first read at 380 ns with a pair of reading beams having a varying relative phase ($\delta_R$).

**Figure 5│ The retrieval efficiency and compensation of Larmor procession in the stored collective spin-wave excitations.** Black circular points show measured retrieval efficiency as a function of storage time when the total phase difference of the total atomic spin wave $\hat{S}_w(z,t)$ is fixed to be in-phase at 380 ns storage time. The time-dependent retrieval efficiency reveals the Rabi oscillation between the two superposition states $\sigma_+|a\rangle$ and $\sigma_-|a\rangle$. Curve a (red curve) is the fitting to the function $R_e = A[1 + \cos(2\Omega_L t - \phi)]e^{-t/t_0}$ (see Methods section), where the fitting parameters are A=0.05, $\Omega_L = 2\pi \times 0.21 MHz$, $\phi = 0.3\pi$ (initial phase), and the lifetime $t_0 \approx 90\mu s$. Curve b (with blue squares) presents the results with compensated Larmor procession by adjusting the relative phase between the two reading beams at different storage times. This indicates that the maximal readout signal intensity can be maintained at any desired retrieving times.

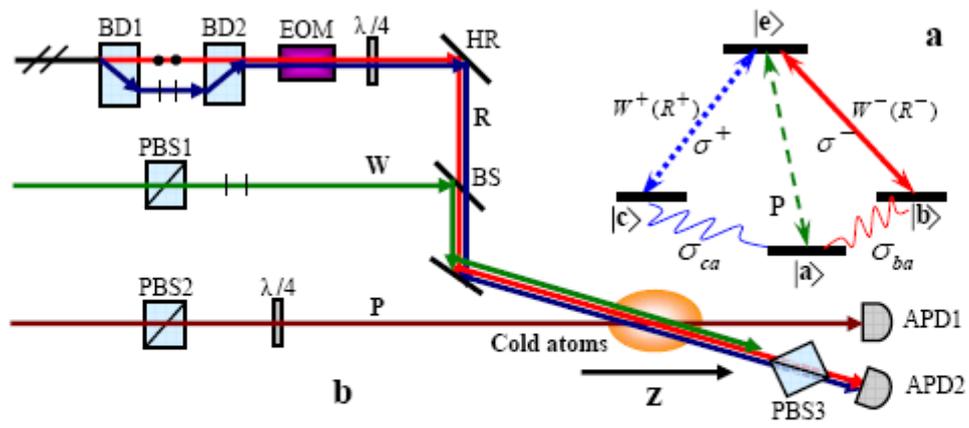

Fig 1

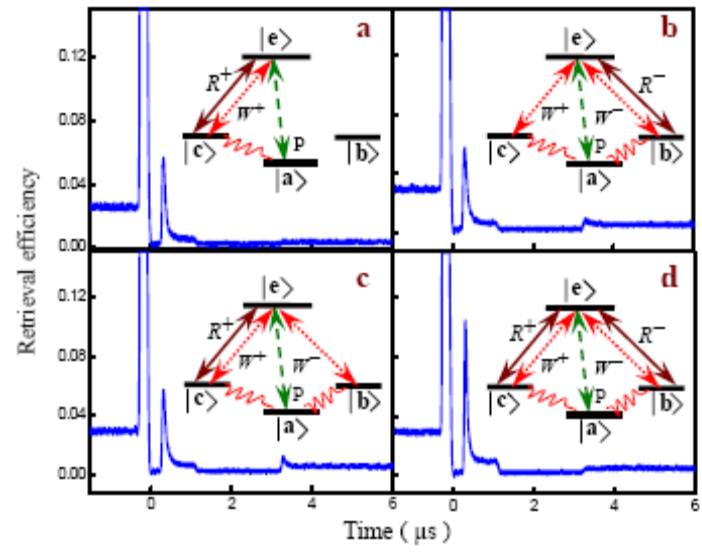

Fig 2

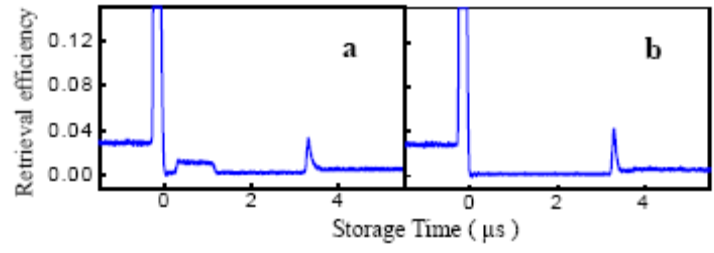

Fig 3

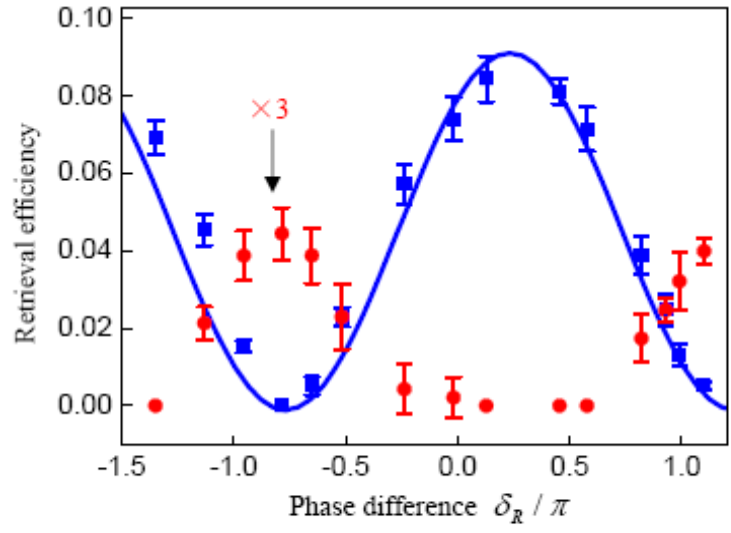

Fig 4

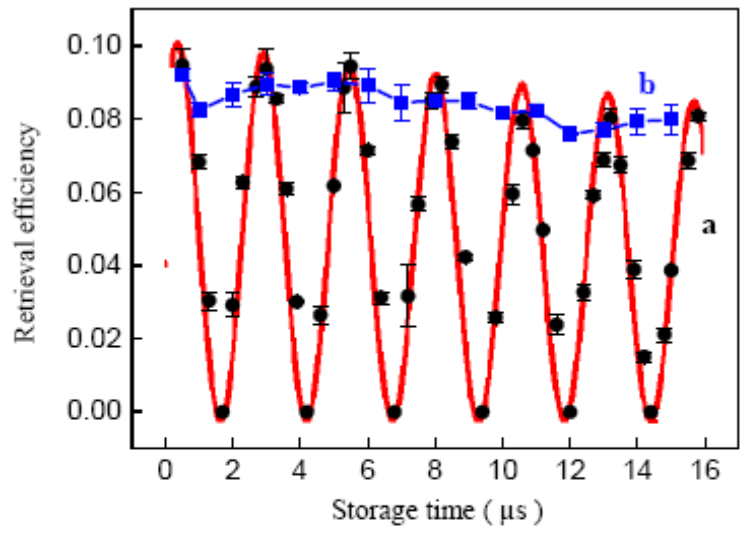

Fig 5